\documentclass{article}
\pdfoutput=1
\usepackage{graphicx}
\usepackage{amsthm}
\usepackage{amsmath} 
\usepackage{amssymb}
\usepackage{graphicx}
\usepackage{float} 
\usepackage{hyperref}
\usepackage{color}
\usepackage{url} 
\usepackage{caption} 
\usepackage{rotating}
\usepackage{multirow}
\usepackage{url}

\begin{document}

\title{A Review of Spatiotemporal Models for Count Data in R Packages. A Case Study of COVID-19 Data}

\author{Mar\'{\i}a Victoria Iba\~nez, \\
\small{mibanez@uji.es}
\and Marina Mart\'{\i}nez-Garcia, \\
\small{martigar@uji.es}
\and Amelia Sim\'o. \\
\small{simo@uji.es} \\
\small{Department of Mathematics-IMAC. Universitat Jaume I. Castelló. Spain}
}

\maketitle
\begin{abstract}

Spatio-temporal models for count data are required in a wide range of scientific fields and they have become particularly crucial nowadays because of their ability to analyse COVID-19-related data.   Models for count data are needed when the variable of interest take only non-negative integer values and these integers arise from counting occurrences.
Several R-packages are currently available to deal with spatiotemporal areal count data. Each package focuses on different models and/or statistical methodologies.
Unfortunately, the results generated by these models are rarely comparable due to differences in notation and methods.
The main objective of this paper is to present a review describing the most important approaches that can be used to model and analyse count data when questions of scientific interest concern both their spatial and their temporal behaviour and we
monitor their performance under the same data set.\\
For this review, we focus on the three R-packages that can be used for this purpose and the different models assessed are representative of the two most widespread methodologies used to analyse spatiotemporal count data: the classical approach (based on Penalised Likelihood or Estimating Equations) and the Bayesian point of view.\\ A case study is analysed as an illustration of these different methodologies.  In this case study, these packages are used to model and predict daily hospitalisations from COVID-19 in 24 health regions within the Valencian Community (Spain), with data  corresponding to the period from 28 June to 13 December 2020. Because of the current urgent need for monitoring and predicting data in the COVID-19 pandemic, this case study is, in itself, of particular importance and can be considered the secondary objective of this work. Satisfactory and promising results have been obtained in this second goal.
\textbf{keyword} {COVID-19; count data; spatiotemporal models; R packages}

\end{abstract}

\section{Introduction} \label{introduccion}
Spatial and temporal models for count data are needed in a variety of settings: agricultural production \cite{besag1999bayesian}, fishing catches \cite{paradinas2017spatio}, volcano eruptions \cite{gusev2008temporal}, crime counts \cite{law2014bayesian}, cases of pulmonary disease \cite{choi2011evaluation}, etc.

In all these examples, the variable of interest take non-negative integer values and these integers arise from counting occurrences of an event in a geographic areal unit in a certain time unit. The observations refer to a set of contiguous non-overlapping areal units for consecutive time periods.  Additionally, a series of covariates are measured for each unit of area and time, these covariates could be common in time or area and they can be discrete, continuous or even factors.

The reasons for modelling these data are diverse and can range from estimating the effect of a risk factor to a response, identifying clusters of contiguous areal units or forecasting future observations. Different modelling strategies have been proposed to deal with this array of scenarios comprising spatiotemporal data. The strategies become more complex when the aim is to build multivariate spatiotemporal models for the joint analysis of different variables that include specific and shared spatial and temporal effects \cite{gomez2019bayesian}.

This kind of data presents two major challenges with respect to classical linear regression models.
Firstly, it is well known that the normal assumption is not appropriate for count data modelling and generalised linear models with Poisson, binomial or negative binomial distributions must be used \cite{nelder1972generalized}.
Secondly,  spatiotemporal autocorrelation, i.e. that observations from geographically close areal units and temporally close time periods tend to have more similar values than units and time periods that are further apart, result in complicated correlation structures, and as a result, parameter estimation is not straightforward and different approaches have been developed for this purpose \cite{anselin1995local,hardisty2010analysing}.

Due to the diversity of applications, data types and conceptual approaches, there is a broad range of literature on spatiotemporal modelling. Two excellent books that provide a gradual entry to the methodological aspects of spatiotemporal statistics and outline some of the standard techniques used in this area are \cite{cressie2015statistics} and \cite{wikle2019spatio}. An overview of different spatiotemporal modelling approaches can also be found in \cite{anderson2017comparison}.

Since December 2019, when the first cases of  the illness caused by the coronavirus SARS-CoV-2 were reported in Wuhan, China, the SARS-CoV-2 virus has spread world-wide. According to the website of the  World Head Organization (WHO) \cite{WHO21}, the virus has caused more than 127 million infections and 2.78 million deaths around the world as of 30 March 2021 and it is currently impossible to predict how many people will be affected by it. The illness caused by the SARS-CoV-2 was given the official name "COVID-19" by the WHO on 11 February 2020 \cite{COVID19name}.

An effective way to control the spread of the infection is to understand and predict key epidemiological data.
Epidemic models have provided powerful insight to study data about the coronavirus pandemic, including the number of new cases and deaths in a given area over time.
Epidemic models to study the spread of infectious diseases date back to the beginning of the twentieth century. The susceptible-infected-recovered (SIR) models developed by Kermack and McKendrick \cite{kermack1927contribution} were the first mathematical models developed to study the transmission dynamics of infectious diseases. The SIR and SEIR (Susceptible, Exposed, Infectious and Removed) models have been improved and used for analysing and characterising the COVID-19 epidemic \cite{fang2020transmission,kucharski2020early,tang2020estimation}, also in Spain  \cite{guirao2020covid, lopez2020modified}. Predicting the future course of epidemics has been another great challenge over time and has become particularly challenging with the rise of new infectious diseases \cite{held2019forecasting}.


In addition to the crucial role played by the above described epidemic models, other models are also being adapted to examine different aspects of the COVID-19 pandemic \cite{fronterre2020covid,dunbar2020endemic}. In particular, mathematical modelling of patient hospitalisation is essential, as it may help raise awareness of a possible collapse of the health-care systems due to an increase in the number of patients needing hospitalisation. Robust prediction models are therefore vital to support decisions on population and community-level interventions to control the spread of the virus and to prevent the collapse of health services.

Models for this type of data are rather different from epidemic models (SIR and SEIR) because the prediction of hospitalisations requires previously obtained COVID-19 data such as the number of people tested and/or infected or the population at risk.
Additionally, models relating to the number of hospitalisations have to manage observations over time in several geographical areas, such as health departments. Each temporal observation relates to an areal unit and, in this case, refers to count measures for the unit: number of COVID-19 hospitalisations in an areal unit per day.
Moreover, given the contribution of people's mobility to the spread of the virus, these models should be able to adjust for people's mobility between neighbouring health areas.

The purpose of the current study was two-fold.
Firstly, to present a review describing three different approaches that can be used to model and analyse count data when the questions of scientific interest concern both their spatial and their temporal behaviour. %

In particular, we revisit three different type of models that are representative of the two most widespread methodologies used to analyse this type of data. The first two models are formulated following the classical statistical paradigm and the last one follows the Bayesian point of view (see \cite{berger2013statistical} for a survey of the hierarchical Bayesian approach and \cite{lawson2018bayesian} for Bayesian disease mapping). Of the two classical approaches one is based on Penalised Likelihood and the other on Estimating Equations \cite{liang1986longitudinal}.

After an exhaustive search among R packages that implement models and procedures to work with spatiotemporal data, we found just three packages that allow us to work with count areal data and we are going to review them.  These three packages are: the \textbf{surveillance} package \cite{meyer2016hhh4}, the model implemented in this package is based on a likelihood model, working on the classical methodology; the \textbf{Mcglm} package \cite{bonat2018multiple}, whose model is based on estimating equations, working on the classical methodology; and finally, the \textbf{CARBayesST} package \cite{lee2018spatio}, based on the Bayesian methodology.

The main properties and characteristics of each of them will be discussed. This can be useful as a guide for scientists in different experimental fields.

We do not claim to provide a comprehensive coverage of all existing methods to deal with count data but to describe and compare the three relevant approaches that are implemented in R packages \cite{R17} to be easily used by researchers.
Other approaches include Dynamic Spatial Panel Data models \cite{elhorst2012dynamic,liesenfeld2017likelihood} that are more usual in the econometric literature; Machine-Learning techniques such as Classification and Regression Trees, Support Vector Machine and Multilayer perceptron Neural Network \cite{martin2020citizen}. Generalized Additive Models have also been used in applied real problems with spatiotemporal data \cite{doi:10.1139/f98-143,Beare2002,smith2019modeling}.  With respect to Bayesian models also different types of spatial, temporal and spatiotemporal random effects, not included in the CARBayesST package, can be used such as non-parametric
estimation of trends \cite{knorr2000bayesian} or splines \cite{ugarte2010spatio,bauer2016bayesian}.

 The second goal of this paper is to apply the reviewed models and compare their performance
 in the prediction of the number of COVID-19 hospitalisations given the number of infected people in the 24 health departments of the  Valencian Community, Spain. The Valencian community is the fourth most populous autonomous community of Spain. It is a rich region with very high residential density along the coast and a lot of tourism and significant exports. Concern about the evolution of the pandemic in this community, led the regional presidency to ask the scientific community for advice about some decision-making regarding the pandemic. Our results will be very useful for this aim.

The article is organised as follows: in Section~\ref{Sec:Methodology} a descriptive analysis of the data is performed, followed by the mathematical details of the three models. Then, the three models are appplied to COVID-19 data in Section~\ref{App}. Our results are discussed in Section~\ref{Discussion} and finally the conclusions are stated in Section~\ref{conclusions}.

\section{Methodology}\label{Sec:Methodology}

\subsection{Dataset}\label{Sec:Dataset}

\begin{figure}[H]
\centering
\includegraphics [scale = 0.65]{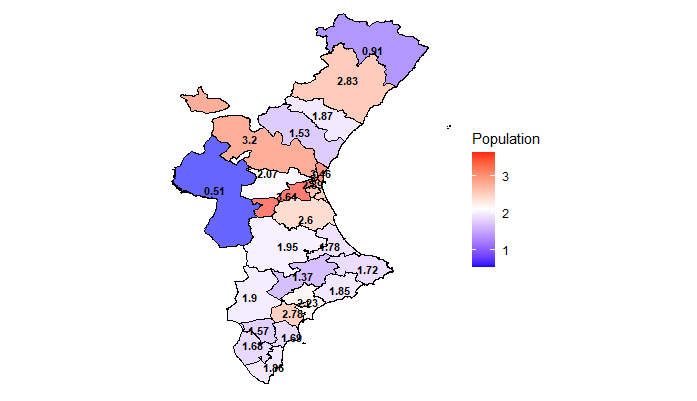} \\
\caption{Location of the Valencian Community within Spain and number of habitants (per 100,000 people) of its 24 health departments.} \label{Fig:Health_regions}
\end{figure}

The data set comprises the number of daily new positive cases of COVID-19 (tested via PCR (polymerase chain reaction) or the antigen test), and the daily number of hospital admissions due to COVID-19 (daily hospitalisations) in the Valencian Community, a large area of Eastern Spain, over 533 days  \textcolor{blue}{(from 28 June 2020 to 13 December 2020)}. This area is organised into $K=24$ different health departments, and the data set comprises both temporal series (hospitalised/new positive cases) for all departments. Figure~\ref{Fig:Health_regions} shows the spatial location of the Valencian Community within Spain and its division into 24 health departments and the distribution of the population ($\times$ 100,000 people) in these 24 health departments. As can be observed, the population in the health departments is very heterogeneous.

The number of new daily positive cases by health department is published regularly on an open data platform of the Generalitat Valenciana (Valencian regional goverment) \cite{datosgva} and the number of daily hospitalisations (by health department) has been provided by the "Data Science for COVID-19 TaskForce" group of the Valencian Community \cite{DSACGV}, with the commitment not to show detailed maps or identifiable information about this variable, which is public only at the aggregated level of the entire Valencian Community. As an illustration, eight of the 24 health departments have been anonymised and  will be used to illustrate all the steps in the different analyses. However, the models are fitted using the data of the 24 health departments and the estimations of the parameters and the goodness of fit measures shown in the paper are related to all of them.

\begin{figure}[H]
\begin{tabular}{ccc}
(a) & (b) \\
\includegraphics[scale=0.5]{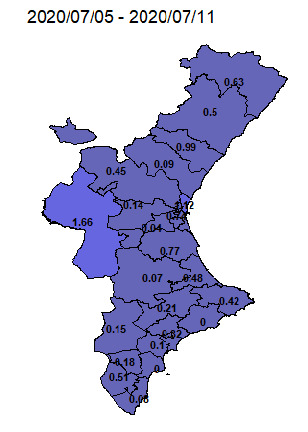}
& \includegraphics[scale=0.5]{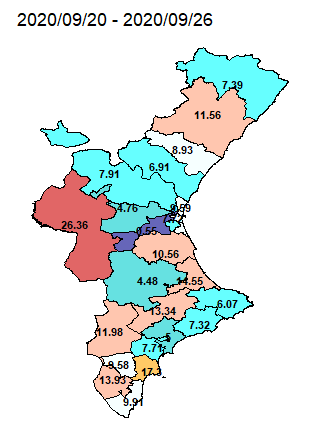}\\
(c) &\\
\includegraphics[scale=0.5]{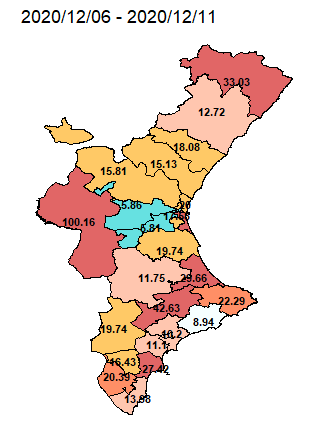}
&\includegraphics[width=20mm]{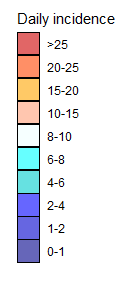}\\
\end{tabular}
\caption{Distribution of mean daily incidence (per 100,000 people) in the 24 health departments of the Valencian Community. The mean daily incidence is computed for one week periods: a)2020/07/05-2020/07/11, b)2020/09/20-2020/09/26 and c)2020/12/06-2021/12/11.}\label{Fig:Positivos_semana}
\end{figure}

Although these data may be subject to temporal biases due to changing testing regimes, among other problems, the mean spatial incidence (number of new cases divided by population size) for three different weeks  (2020/07/05-2020/07/11, 2020/09/20-2020/09/26 and 2020/12/04-2020/12/11) plotted in Figure~\ref{Fig:Positivos_semana}, shows strong variation across the different health departments over time.

\begin{figure}[H]
\begin{tabular}{cc}
(a) & (b)\\
\includegraphics[scale=0.35]{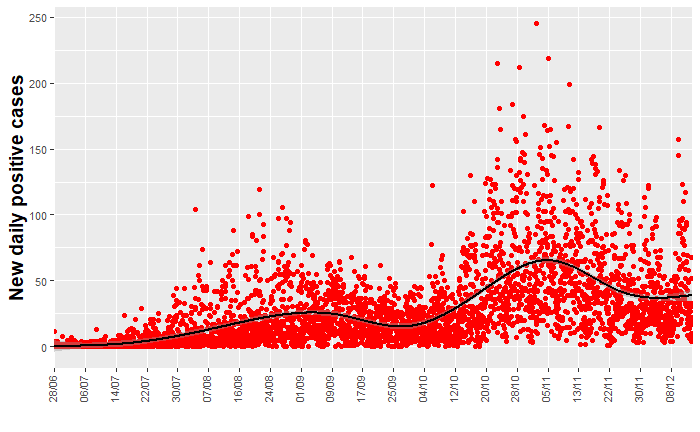}
& \includegraphics[scale=0.35]{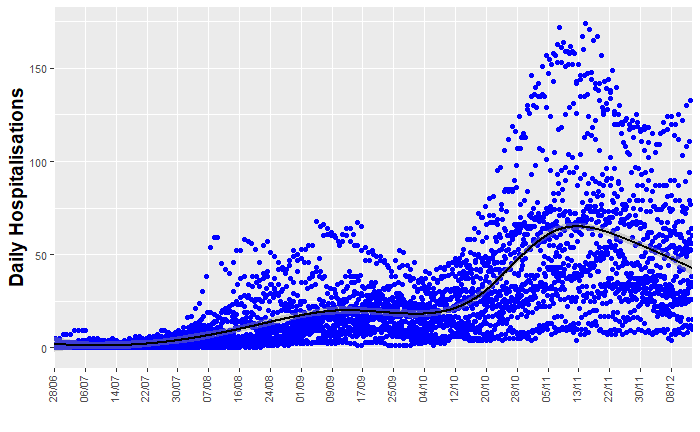} \\
(c) & (d)\\
\includegraphics[scale=0.35]{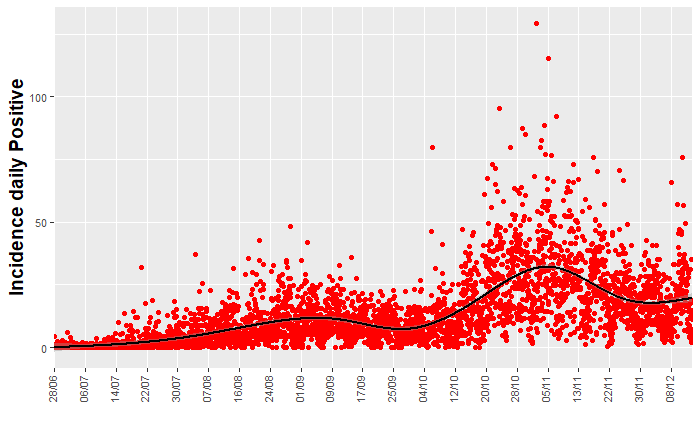}
& \includegraphics[scale=0.35]{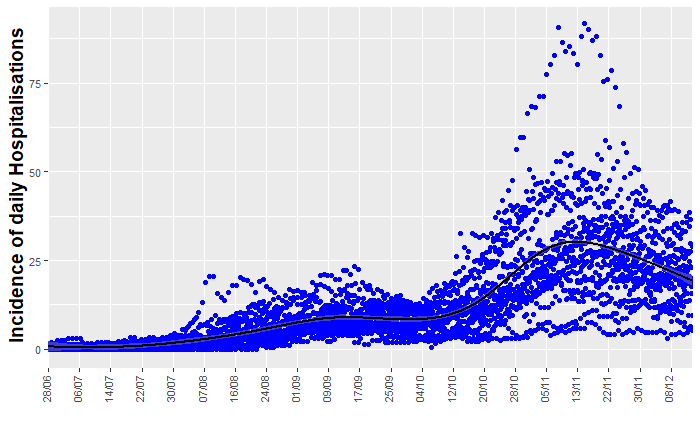} \\
(e) & \\
\includegraphics[width=0.36\textwidth]{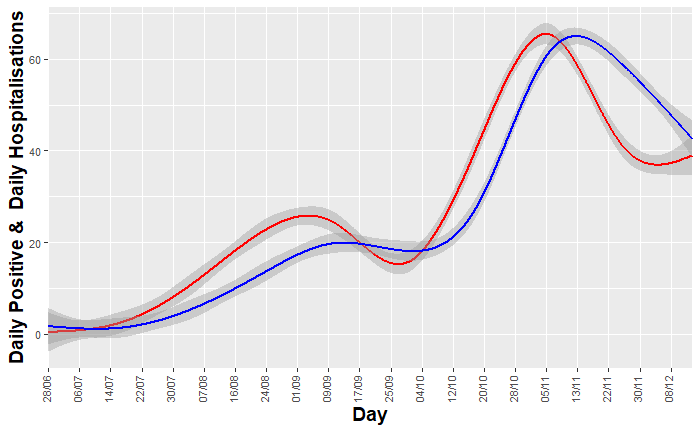}
&\\
\end{tabular}
\caption{Temporal trend of COVID-19 for (a) new daily positive cases; (b) daily hospitalisations,
per health department, where each point represents the data of one of the 24 health departments;(c) incidence of new daily positive cases, positives divided by region population and multiplied by 100,000; (d)  incidence of daily hospitalisations; e) temporal trend of COVID-19 for the daily sum of the 24 health departments of both time series: daily positive cases (red) and daily hospitalisations (blue) with a mean confidence interval of 95\% (grey).
}\label{Fig:DescriptivoGlobal}
\end{figure}

Figure~\ref{Fig:DescriptivoGlobal} a) and b) show time series plots for both variables of interest: the number of daily new positive cases and the number of daily hospitalisations per health department. From now on, daily hospitalisations we will mean the total number of people admitted to hospital due to COVID-19 each day.  Figure~\ref{Fig:DescriptivoGlobal} a) shows a peak of new cases in late August, and another in mid-November, two and a half months apart, and the same temporal pattern appears in the series of hospitalisations.

As different health department have different population sizes, Figures~\ref{Fig:DescriptivoGlobal} c) and d) show the relative data, it is said, the number of new positive cases and daily hospitalisations corrected by the population size of each health department (incidence values). As can be seen, temporal patters remain unchanged.
Figure~\ref{Fig:DescriptivoGlobal} e) shows the total number of new positive cases and hospitalisations per day (adding the values of the 24 health departments). From Figure~\ref{Fig:DescriptivoGlobal} e) it can be seen that there is a time lag between both time series.

\begin{figure}[h!]
\centering
\includegraphics [height = 6cm]{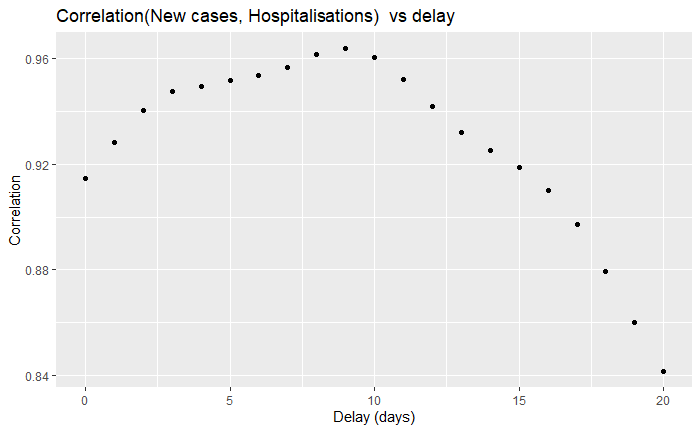}
\caption{Cross-correlogram, between daily new positives and daily hospitalisations. It shows the Pearson correlation between both series as a function of the displacement (days) of daily positives relative to the daily hospitalisations.}\label{Fig:Correlation}
\end{figure}

To estimate the delay between both time series a cross-correlogram (see Figure~\ref{Fig:Correlation}) has been used. It plots a measure of correlation of both time series (in this case Pearson correlation), as a function of the displacement (days) of daily positives relative to the daily hospitalisations. Figure~\ref{Fig:Correlation} shows that there are two maximums in the correlation  at the time lag of 9 and 5 days.

\begin{figure}[h!]
\begin{tabular}{cc}
a)& b)\\
\includegraphics[scale=0.35]{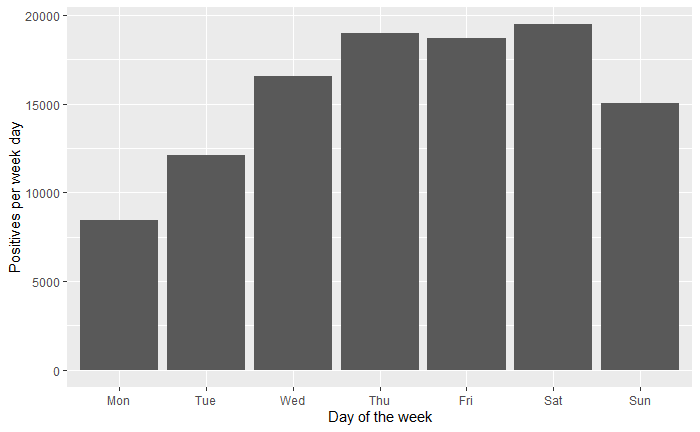}&
\includegraphics[scale=0.35]{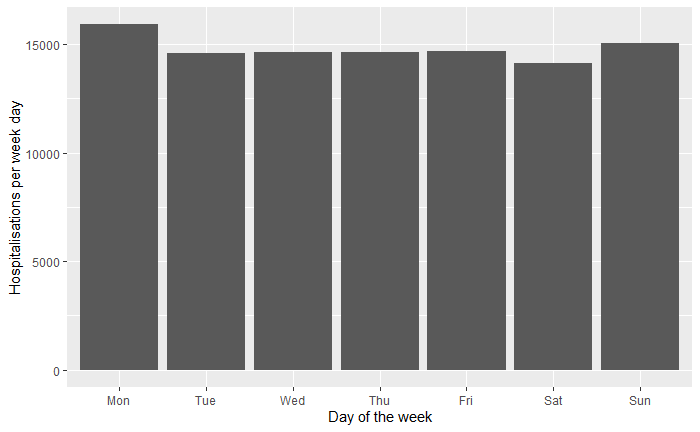}\\
(c)& \\
\includegraphics[scale=0.35]{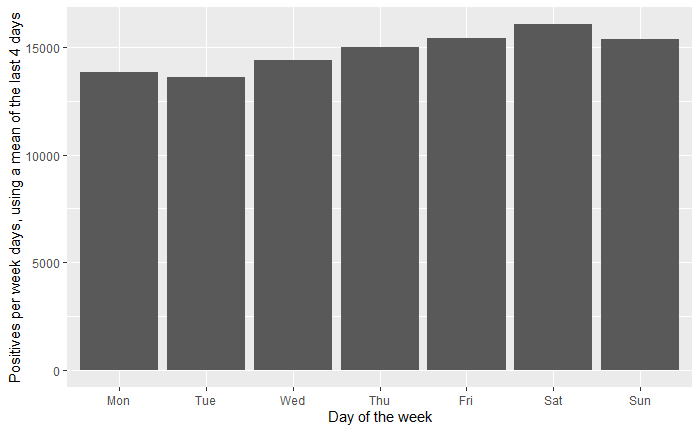}
\end{tabular}
\caption{Data dependency with respect to the day of the week: a)Daily positives vs day of the week; b) Daily hospitalisations vs day of the week; c) Mean of daily positives in the last 4 days vs day of the week. }\label{Fig:Pos_week_day}
\end{figure}

These data may be subject to temporal biases due to under-reporting at weekends and/or on non-working days. Figure~\ref{Fig:Pos_week_day} a) shows a great variability in the number of positives depending on the day of the week. As can be seen in Figure~\ref{Fig:Pos_week_day} b), this effect does not hold for the number of daily hospitalisations. Therefore, this effect is an artefact, due to when the official data is reported rather than a real effect of the virus. To minimise this effect, henceforth, we will take the mean of the last 4 days as the daily hospitalisations. By doing this smoothing, we reduce this effect, as can be seen from Figure~\ref{Fig:Pos_week_day} c).

\begin{figure}[h!]
\begin{tabular}{c}
\includegraphics [height = 10cm]{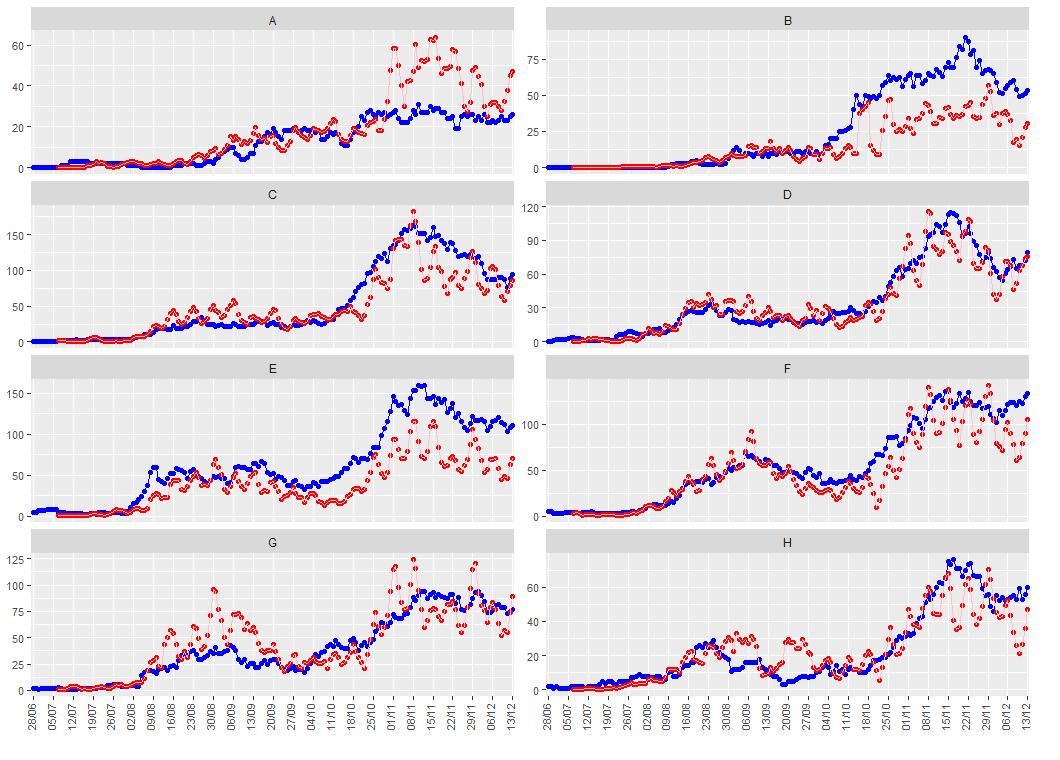}
\end{tabular}
\caption{Temporal trend of COVID-19 for daily positives (red line) and daily hospitalisations (blue line) in the eight health departments used as an illustration, with a time lag of 9 days in the daily positive cases, and a smoothing of 4 days.} \label{Fig:GeneralDelay}
\end{figure}

Finally, both effects are corrected. Figure~\ref{Fig:GeneralDelay} shows the smoothed number of positive daily cases (in red) together with the number of people hospitalised due to COVID-19 (in blue) for the eight illustrative health departments, with a time delay of 9 days between both series.

The spatial adjacency matrix between these departments is shown in table \ref{tabla:vecindad}. In this table, values equal to 1 signal neighbouring regions, i.e. those that share a geographical border.

\begin{table}[H]
\caption{Spatial adjacency matrix of the 8 health departments used for illustration purposes. If two health departments,$i,j$, are neighbours, the matrix value is one, otherwise 0.}\label{tabla:vecindad}
\begin{tabular}{c|cccccccc}
\hline
 & A & B & C & D & E & F & G& H\\
\hline
A&0 & 1 & 0 & 0&  0 & 0  & 0   &0\\
B&1 & 0 & 1 & 1&  0 & 0  & 0   &0\\
C&0 & 1 & 0 & 1&  1 & 1  & 1   &0\\
D&0 & 1 & 1 & 0&  1 & 1  & 1   &1\\
E&0 & 0 & 1 & 1 & 0 & 1  & 1   &1\\
F& 0&  0 & 1 & 1 & 1&  0 &  1  & 1\\
G&  0&  0 & 1&  1 & 1&  1 &  0 &  0\\
H&  0&  0  &0&  1  &1&  1  & 0 &  0\\
\hline
\end{tabular}
\end{table}

\begin{figure}
\includegraphics[scale=0.6]{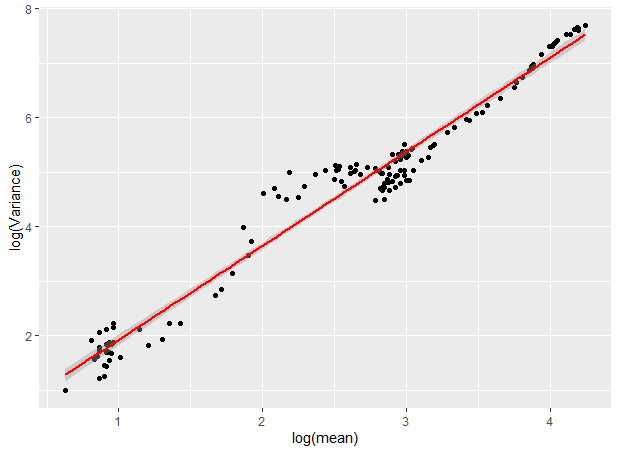}
\caption{Mean vs variance of the number of hospitalisations per day of the 24 health departments for each of the 533 days studied. The data are represented on a log-log scale (black dots). The red line, indicates a smoothing of the data tendency.} \label{Fig:mediavarianza}
\end{figure}

To conclude, when looking for the best model, we explored the relationship between the mean and the variance of the hospitalisation data collected at each instant of time. Figure~\ref{Fig:mediavarianza} shows that there is a potential relationship between mean and variance. This relationship is crucial for the modelling, as will be seen in the following sections.

\subsection{Models} \label{Sec:ModelsandMethods}
Throughout this section $Y_{kt}$  will denote the observation taken in the $k-$th areal unit $S_k$ at time $t$, for $k = 1, \ldots ,K$ and $t=1,\cdots, N$.
Then, $\textbf{Y}_t=(Y_{1t},\ldots,Y_{Kt})$, ($t = 1, \ldots ,N$), will be a spatiotemporal count series, i.e. count data recorded in the areal units for consecutive discrete time periods.
    We assume that we also have space-time varying covariates $\textbf{X}_t=(X_{1t},\ldots,X_{Kt})$ recorded at the same times and locations. Our main objective will be to predict future observations of the spatiotemporal time series $\textbf{Y}_t$, by taking into account the spatiotemporal covariates $\textbf{X}_t$ and the spatial and temporal relationships between the observations.

\subsubsection{Endemic-epidemic models. R package surveillance \cite{meyer2016hhh4}}\label{Package:surveillance}

Endemic-epidemic (EE) models are a class of statistical time series models for
multivariate surveillance counts proposed by \cite{held2005statistical} and
extended in \cite{paul2011predictive} and \cite{meyer2014power, meyer2016hhh4}).

In its current formulation and implementation in the R package surveillance \cite{meyer2016hhh4}, the EE framework uses incidence from the preceding week, $t-1$, to explain the incidence in week $t$. So, the counts,  $Y_{kt}|\textbf{Y}_{t-1}$, are assumed to be Poisson or Negative Binomial distributed with the conditional mean:

\begin{equation}\label{eq:hhh4}
\mu_{kt}=e_{kt}\nu_{kt} +\lambda_k Y_{k, t-1}+\phi_k \sum_{q\neq k} w_{qk} Y_{q,t-1}, \ \ \ \ \nu_{kt}, \lambda_k, \phi_k >0,
\end{equation}
and overdispersion parameter, in the Negative Binomial case, $\psi_{k}>0$.

The first component of the summation is called the endemic component and captures information not directly linked to observed cases from
the previous day. This component can cover exogenous factors such as temporal trends, seasonality, sociodemographics, and/or population. As an example, in spatial applications, $e_{kt}$ can refer to the fraction of the population living in region $k$ at time $t$. The remaining terms in Eq. \ref{eq:hhh4} constitute the epidemic component and describe how the incidence
in region $k$ is linked to previous cases in the same and adjacent regions. The two terms of this epidemic component are usually denoted as  'autoregressive' and 'spatiotemporal' component, respectively.

The parameters $\nu_{kt}$, $\lambda_k$ and $\phi_k$ are constrained
to be non-negative and can be modelled by allowing for log-linear
predictors in all three components, as sine-cosine terms to account for seasonality
\cite{held2012modeling}, long-term temporal trends or/and covariates
\cite{bauer2018stratified, cheng2016analysis}.

\begin{eqnarray}
  log(\nu_{kt})      &=& \alpha^{(\nu)}+b_k^{(\nu)}+\beta^{(\nu)}z_{kt}^{(\nu)} \label{Eq:hhh4covariates}\\
  log(\lambda_k)      &=& \alpha^{(\lambda)}+b_k^{(\lambda)}+\beta^{(\lambda)}z_{kt}^{(\lambda)}\nonumber \\
  log(\phi_k)      &=& \alpha^{(\phi)}+b_k^{(\phi)}+\beta^{(\phi)}z_{kt}^{(\phi)}.\nonumber
\end{eqnarray}

This form allows for fixed intercepts $\alpha^{(.)}$, region-specific intercepts $b_k^{(.)}$ and exogenous
covariates $z_{kt}^{(.)}$  in each model compartment. Population fraction, population density, border effects, etc.
can be used as covariates. The
region-specific intercepts $b_k^{(.)}$, can be treated as fixed effects or as random effects accounting for heterogeneity between the regions. When they are treated as random effects, they
 are assumed to be independent and identically distributed across $k$, but can be correlated across the
model components, following a Gaussian distribution:
\begin{eqnarray*}
b_k := (b_k^{(\nu)}, b_k^{(\lambda)}, b_k^{(\phi)}) \sim N((0,0,0), \Sigma_b).
\end{eqnarray*}
We will see this part in more detail in Section \ref{App1:EE}.

Maximum likelihood (ML) estimates are obtained using penalised likelihood approaches.

This basic model has been extended to cover other different aspects of disease modelling (see \cite{dunbar2020endemic} for references). Recent extensions include 
methodology to adjust for under-reporting \cite{dunbar2020endemic,hhh4underreporting} or to allow different lags in the auto-regressive part of the model (package \textbf{hhh4addon} \cite{hhh4addon, bracher2017periodically}), modelling the conditional mean $Y_{kt}|\textbf{Y}_{t-1},\cdots \textbf{Y}_{t-D}$ as:

\begin{equation}\label{eq:hhh4addon}
\mu_{kt}=e_{kt}\nu_{kt} +\lambda_k \sum_{q}\sum_{d=1}^D w_{qk} Y_{q,t-d}, \ \ \ \ \nu_{kt}, \lambda_k >0,
\end{equation}
where $D$ is the maximum lag considered.

In \cite{meyer2017incorporating}, the authors extend the basic endemic-epidemic spatiotemporal model to fit multivariate time series of counts $y_{gkt}$ stratified by (age) groups in addition to spatial regions. They therefore define a contact matrix $C=(c_{g'g})$, where $c_{g'g}\geq 0$ quantifies the average number of contacts of an individual of group $g'$ with individuals of group $g$, and the spatiotemporal model is now modelled as:

\begin{equation}\label{eq:hhh4contacts}
\mu_{gkt}=e_{gkt}\nu_{gkt} +\lambda_{gkt} \sum_{g'q} c_{g'g}w_{qk} Y_{g',q,t-1}, \ \ \ \ \nu_{kt}, \lambda_k >0,
\end{equation}
where both the endemic and epidemic predictors may gain group-specific effects. This model is implemented in the R-package \textbf{hhh4contacts} \cite{hhh4contacts}.

\vspace*{0.2cm}
\emph{Forecast}

The \textbf{surveillance} package uses the function \textbf{hhh4} to fit the models and implements the \textbf{oneStepAhead()} function, which computes successive one-step-ahead predictions for the fitted model, also providing confident intervals for the predictions and plot methods.
The associated scores-method computes a number of (strictly) proper scoring rules based on such one-step-ahead predictions; see \cite{paul2011predictive} for details.

A discussion of suitable measures to evaluate the quality of a point forecast can be found in \cite{gneiting2011making} and  several scoring rules based on the one-step-ahead predictions \cite{paul2011predictive}  are implemented in the function \textbf{scores}, although  we will consider the root mean squared error of the predictions (RMSEp). Another function implemented in the package related to the \textbf{oneStepAhead()} function  is the \textbf{calibrationTest} function, which implements calibration tests for Poisson or Negative Binomial predictions of count data based on proper scoring rules; it is described in detail in \cite{wei2014calibration}.

Long-term predictions do not have much sense in our context because we do not know the long-term evolution of the covariates.

\subsubsection{Multivariate covariance generalised linear models. R package Mcglm} \label{Sec:MCGLM}

Under the same previous assumption of predicting  $\textbf{Y}_t=(Y_{1t},\ldots,Y_{Kt})$ in terms of spatiotemporal correlations and $\textbf{X}_t=(X_{1t},\ldots,X_{Kt})$ covariates,  we can use the multivariate covariance generalised linear model (McGLM) introduced in \cite{bonat2016multivariate}.
This model is a general and flexible statistical model to deal with multivariate count data that explicitly models the marginal covariance matrix combining a covariance link function and a matrix linear predictor composed of known matrices.

Let $\textbf{Y}_{K\times N} = \{\textbf{Y}_1, \cdots , \textbf{Y}_N \}$ be the outcome matrix and let $M_{K\times N} = \{\mu_1, \cdots ,\mu_N\}$ denote the corresponding matrix  of expected values.

Given $\Sigma_t$ the $K\times K$ covariance matrix within the response variable $Y_t$ for $t = 1, \cdots,N$ and $\Sigma_b$ the $N \times N$ correlation matrix whose components denote the correlation between outcomes, the  McGLM as proposed by \cite{bonat2016multivariate} is given by
\begin{eqnarray}\label{eq:mcglm1}
E(\textbf{Y})&=&(\mu_1, \cdots ,\mu_N) =(g_1^{-1}(X_{1}\beta_1),...,g_N^{-1}(X_{N}\beta_N))\\
Var(\textbf{Y})&=&C=\Sigma_t \bigotimes_G \Sigma_b, \nonumber
\end{eqnarray}
where $g_t$ are monotonic differentiable link functions, $X_t$ denotes an $K\times k_t$ design matrix, $\beta_t$ is a regression parameter vector to be estimated, and

 \begin{eqnarray*}
 \Sigma_N\bigotimes_G \Sigma_b=Bdiag(\tilde{\Sigma}_1,\cdots ,\tilde{\Sigma}_N)(\Sigma_b \bigotimes I)Bdiag(\tilde{\Sigma}_1^{-T},\cdots,\tilde{\Sigma}_N^{-T})
 \end{eqnarray*}
is the generalised Kronecker product \cite{martinez2013general}. The matrix $\tilde{\Sigma}_t$ denotes the lower triangular matrix of the Cholesky decomposition of $\Sigma_t$. The operator $Bdiag$ denotes a block diagonal matrix, and $I$ denotes a $K \times K$ identity matrix.

A key point for specifying McGLMs for mixed types of outcomes is the specification of the covariance matrix within outcomes $\Sigma_k$. Following \cite{bonat2016multivariate}, we define the covariance within outcomes by
\begin{eqnarray}
\Sigma_t = V(\mu_t; p_t)^{\frac{1}{2}}(\Omega(\tau_t))V(\mu_t; p_t)^{\frac{1}{2}},
\end{eqnarray}
For modelling count outcomes, they propose to adopt the  Poisson-Tweedie dispersion function  \cite{jorgensen2016discrete} so that
\begin{eqnarray}
V(\mu_t; p_t)=diag(\mu_t^{p_t})
\end{eqnarray}
is a diagonal matrix whose main entries are given by the power variance function. The Poisson-Tweedie family of distributions provides a rich class of models to deal with count outcomes, since many important distributions appear as special cases; examples include the Hermite ($p$ = 0), Neyman Type A ($p$ = 1), Negative Binomial (p = 2) and Poisson-inverse Gaussian ($p$ = 3).

The dispersion matrix $\Omega(\tau_t)$ describes the part of the covariance within outcomes that does not depend on the mean structure. Jorgensen et al. \cite{jorgensen2016discrete}, among others, propose to model the dispersion matrix using a matrix linear predictor combined with a covariance link function, i.e.

\begin{equation}
h(\Omega(\tau_t))=\tau_{t0}Z_{t0}+\cdots+\tau_{tD}Z_{tD},
\end{equation}
where $h$ is the covariance link function, $Z_{td}$ with $d = 0,\cdots , D$ are known matrices reflecting the covariance structure within the response variable $Y_t$, and $\tau_r = (\tau_{r0}, \cdots , \tau_{rD})$ is a $(D+1)\times 1$ vector of dispersion parameters.

McGLMs are fitted based on the estimating function approach described in detail by \cite{bonat2016multivariate}and \cite{jorgensen2004parameter}. A general overview of the algorithm and the asymptotic distribution of the estimating function estimators can be found in \cite{bonat2018multiple}.  As a method for selecting the components of the matrix linear predictor (variable selection), the score information criterion (SIC) is proposed. This is an important tool to assist with the selection of the linear and matrix linear predictor components, but unfortunately it is less useful for comparing models fitted using different link, variance or covariance functions.

\vspace*{0.2cm}
\emph{Forecasting}

Unfortunately, the \textbf{mcglm} package does not have any function implemented to predict future observations. If we do not know to implement a more sophisticated routine, once the model has been estimated, the \textbf{mc link} function can be used. This function returns the inverse of the link function applied to the linear predictor i.e. $\mu = g^{-1}(X\beta)$, as an approximation of the predictions sought.

\subsubsection{Bayesian hierarchical generalised linear models. CARst package } \label{secBayesian}
A great variety of spatiotemporal models for count data using generalised linear models (GLM) can be found in the Bayesian literature. To model these data a hierarchical model with spatiotemporal structured prior distributions
is used. The spatiotemporal structure is modelled via sets of autocorrelated random effects with conditional autoregressive and its spatiotemporal extensions priors. An excellent review can be found in \cite{lee2018spatio}.

These methods have had a remarkable development, especially in disease mapping, thanks to the availability of estimation methods based on Monte Carlo Markov Chain.
With respect to the software packages for implementing these models, although a great quantity of software packages can be found for implementing purely spatial models, such as BUGS \cite{lunn2009bugs} and R-INLA \cite{rue2009approximate}, software for spatiotemporal modelling is much less well developed and mainly focuses on geostatistical data. This was the motivation for developing the CARBayesST R package \cite{lee2018spatio}. This package can fit several models for count data with different spatiotemporal structures. A useful tutorial is provided by \cite{lee2020tutorial}.
CARBayesST package has been recently used to study the case-fatality risk by COVID-19 in Colombia \cite{polo2020bayesian}.

The general Bayesian hierarchical model for spatiotemporal count data is as following:
\begin{eqnarray}
  Y_{kt}      &\sim& f(y_{kt}|\mu_{kt},\nu) \\ \label{eq:hierarchical1}
  g(\mu_{kt}) &=&  X_{kt}\beta +\psi_{kt} \\ \label{eq:hierarchical2}
  \beta      &\sim& N(\mu_{\alpha}, \Sigma_{\beta}).\label{eq:hierarchical3}
\end{eqnarray}

The probability function $f$ is in the exponential family (not necessarily a Gaussian distribution), $\beta$ is the vector of covariate regression parameters, and a multivariate Gaussian prior is assumed. $g$ can be any monotonic differentiable link function, and  $\psi_{kt}$ is a latent component for areal unit $k$ and time period $t$ encompassing one or more sets of spatiotemporally autocorrelated random effects, we denote $\psi_t=(\psi_{1t},\ldots,\psi_{Kt})$.

In this paper, we are just focusing on the models implemented in the CARBayesST package. In this package, binomial, Gaussian and Poisson data
models can be used for the first level of the model, $f$ in Eq. \ref{eq:hierarchical1}, and different spatiotemporal structures for $\psi_{kt}$ in Eq.\ref{eq:hierarchical2} are given.

All models implemented in this package use random effects to introduce spatial autocorrelation into the response variable. For this purpose, CAR-type prior distributions and their space-time extensions are used. Spatial autocorrelation is induced via a non-negative symmetric matrix of adjacency
$W= (w_{kj})$, where $w_{kj}$ represents the spatial closeness between units $(S_k, S_j)$.
Larger valued elements represent spatial closeness between the two areas in question and spatially autocorrelated random effects, whereas zero values correspond to areas that are not spatially close and conditionally independent random effects given the remaining.

The models are outlined in Table~\ref{TablaBayesianModels} \cite{lee2018spatio}.
In all cases, inference is based on Markov chain Monte Carlo (MCMC) simulation.

\begin{table}[H]
\caption{Summary of the models available in the CARBayesST package. \label{TablaBayesianModels}}
\begin{tabular}{|l|l|}
\hline
ST.CARlinear \cite{bernardinelli1995bayesian}&
Spatially varying linear time trends\\
& model \\
\hline
 ST.CARanova  \cite{knorr2000bayesian}
& Spatial and temporal autoregressive main  \\
& effects and  independent interaction model \\
\hline
ST.CARsepspatial  \cite{napier2016model}
&
Common temporal trend but varying spatial \\
& surfaces model \\
\hline
 ST.CARar \cite{rushworth2014spatio}&
Spatially autocorrelated autoregressive of \\
& order 1 time series model \\
\hline
ST.CARadaptive \cite{rushworth2017adaptive}& Spatially adaptive smoothing model for \\
&localised spatial smoothing \\
\hline
ST.CARlocalised \cite{lee2016quantifying}&
 spatiotemporal  clustering model\\
\hline
\end{tabular}
\end{table}

\vspace*{0.2cm}
\emph{Forecasting}

    In order to predict future observations of the response variable, simulations from  the posterior predictive distribution (the distribution of possible unobserved values conditional on the observed values) have to be obtained.

This predictive density can be approximated by Monte Carlo integration. If we denote the vector with all the parameters of the model  by $\theta$ :

$$ f(\textbf{Y}_{N+h}|(\textbf{Y}_1,\ldots,\textbf{Y}_N))\approx \frac{1}{n}\sum_j^n f(\hat{\textbf{Y}}_{N+h}|\theta^{(j)}, (\textbf{Y}_1,\ldots,\textbf{Y}_N)). $$

If a representative value is wanted, the mean of the predictive density can be obtained by taking into account the property that $E_{\theta} E(\textbf{Y}_{N+h}\mid \theta) )$ and approximating the mean again with respect to $\theta$:

$$E(\textbf{Y}_{N+h}|(\textbf{Y}_1,\ldots,\textbf{Y}_N))=\frac{1}{n}\sum_j^n E(\textbf{Y}_{N+h}|\theta^{(j)},(\textbf{Y}_1,\ldots,\textbf{Y}_N))$$

Unfortunately, the CARBayesST package does not have any function implemented to predict future observations. for $h$=1 it can be easily implemented using the samples of the posterior distributions of the parameters. For $h>1$ the simulation of the posterior distribution of the random effects should be implemented. In this case, for users who are not experts in R programming, just an approximation can be obtained, approximating the value of the random effect by the that of the previous prediction.
This approximation will be used in Section \ref{Appl:Bayes}.

\section{Application}\label{App}

In this section, all the models reviewed are applied to the COVID-19 data described in Section \ref{Sec:Dataset}. Continuing with the notation introduced,  $\textbf{Y}_t=(Y_{1t},\ldots,Y_{Kt})$ refers to the observed hospitalisation data at time $t$ in the $K=24$ health departments of the Valencian Community, $\textbf{X}_t=(X_{1t},\ldots,X_{Kt})$ refers to the positive cases at time $t-lag$, where $lag$ is the previously determined and $p_k$ is the population at risk in the $k$-th health department.

Since our primary goal is forecasting, the mean squared errors of the predictions  (RMSEp) up to a five-day horizon are calculated. This is a classical approach to measure their performance. Therefore, data from 28 of June 2020 to 8 of December 2020 are used to fit the different models. The root mean square error comparing observed and fitted values  (RMSEf) are computed in all cases to describe the goodness the fits. Data from 9 to 13 of December 2020, jointly with the five-days horizon forecasts of each  model, are used to compute the root mean square error of the predictions (RMSEp). If we want to use real data of positive cases, the horizon of prediction is limited, but taking a horizon of five is considered enough, within our possibilities, to prevent the collapse of hospitals.

Each model is adjusted using the corresponding statistical methodology included in its corresponding R package. Additional specific measures of goodness of fit are given; these other measures will be useful to compare different models within the same R package.

\subsection{Endemic-epidemic models}\label{App1:EE}


As stated in Section \ref{Package:surveillance}, let us consider that the counts of daily hospitalisations in the $k-$th health department, on the $t-$th day, $Y_{kt}$, follows a Poisson distribution with mean  as in Eq.\ref{eq:hhh4}.

If $p_k$ denotes the population of the $k-$th health department, we assume in Eq.  \ref{eq:hhh4} known population fractions
\begin{eqnarray*}
e_{kt}=e_k=\frac{p_k}{\sum_{k=1}^{24}p_k}, \hbox{     } \forall t,
\end{eqnarray*}
and that $w_{qk}=I(q \sim k)$, i.e, $w_{qk}=1$ if both health areas have a common geographic border (assuming the epidemic only arrives from adjacent health areas) and 0 otherwise. Weights $w_{qk}\! = \frac{w_{qk}}{\sum_q w_{qk}}$ are normalised and restricted to be positive.

Covariates such as number of positive cases can be added to the model in different ways \cite{herzog2011heterogeneity,meyer2016hhh4}.
The simplest way is to include the covariates $x_{kt}$ in  the formulation in the endemic part of the model, for example considering:

\begin{itemize}
\item Model 1:
\begin{eqnarray}\label{hhh4Model0}
\mu_{kt}&=& e_{k}\nu_{kt} +\lambda_{kt} Y_{k, t-1}+\phi_{kt} \sum_{q\neq k} w_{qk} Y_{q,t-1}, \ \ \ \ \nu_{kt}, \lambda_k, \phi_k >0,\nonumber\\
log(\nu_{kt})   &=& \alpha^{\nu}+\beta_1^{\nu} x_{kt}, \hbox{     }\forall k \nonumber \\
log(\lambda_{kt})   &=& \alpha^{\lambda}, \hbox{     }\forall t,k  \nonumber \\
log(\phi_{kt})     &=& \alpha^{\phi}, \forall k,t \nonumber\\
w_{qk} &=&  I(q \sim k).
\end{eqnarray}
\end{itemize}

We are going to consider two possibilities regarding the covariates. Case 1: consider the smoothed number of new positive cases at a lag 9 as a covariate. Case 2: consider the smoothed number of positive new cases at a lag 9 and the smoothed number of new positive cases at a lag 5 as covariates. Many works include a seasonal effect in the model of the parameters $\log(\nu_{kt}), log(\lambda_{k})$ and/or $log(\phi_{k})$, but we would expect this seasonal effect to be included in the covariates (time series of positive cases), so we include it only in the model of $log(\lambda_{k})$ and $log(\phi_{k})$.

All unknown parameters  are estimated directly by maximising the corresponding log-likelihoods using numerical optimisation routines (see \cite{paul2008multivariate}). The estimates obtained and several goodness of fit measures for this model are shown in Table~ \ref{tabla:parametrosmodelos1}.

Having estimated the parameters of the model, the fitted mean can be compared with the observed counts in order to check the goodness of fit, but additionally, we can see the contribution to this fitted mean of the endemic and autoregressive components. The average of the proportions of the mean explained by the different components are also shown in Table~\ref{tabla:parametrosmodelos1}. Note that the proportion explained by the epidemic component is around  97$\%$ in both cases, being by far the component with the greatest  influence on the value of the total fit. So, there is a high influence of the within-health area autoregressive component, with very little contribution of adjacent areas and a rather small endemic incidence.

\begin{table}[H]
\caption{Estimations, goodness of fit measures and contribution of the endemic and autoregressive components to the global fit. RMSEf is the RMSE of fitted values and RMSEp the RMSE of predictions.\label{tabla:parametrosmodelos1}}
\begin{tabular}{|c|cc|cc|cc|cc|}
\hline
&\multicolumn{2}{c|}{Model 1.1} & \multicolumn{2}{c|}{Model 1.2}  \\
\hline
& Estimate & Std. Error & Estimate & Std. Error \\
\hline
$\alpha^{\lambda}$& 0.985 & 0.006  & 0.985 &  0.006\\
$\alpha^{\phi}$   & 0.002 &  0.0008&0.002 & 0.0008\\
$\alpha^{\nu}$    & 4.061 &  0.709 &4.085 & 0.7117 \\
$\beta_1$ &   1.019 &   0.003 & 1.005 & 0.019\\
$\beta_2$ &   - &   - &1.014 &  0.0153 \\
\hline
Log-likelihood: & \multicolumn{2}{c|}{-8559.39} &\multicolumn{2}{c|}{-8558.74}\\
AIC: &  \multicolumn{2}{c|}{17126.78} & \multicolumn{2}{c|}{17127.48} \\
BIC: &  \multicolumn{2}{c|}{17151.64} & \multicolumn{2}{c|}{17158.56} \\
\hline
RMSEf & \multicolumn{2}{c|}{0.54}&  \multicolumn{2}{c|}{0.53}\\
RMSEp & \multicolumn{2}{c|}{6.13}&  \multicolumn{2}{c|}{6.18}\\
\hline
endemic &  \multicolumn{2}{c|}{1.39 $\%$ }& \multicolumn{2}{c|}{1.47 $\%$ } \\
epi.own &  \multicolumn{2}{c|}{97.28 $\%$}  & \multicolumn{2}{c|}{97.21$\%$ }\\
epi.neigbours &  \multicolumn{2}{c|}{1.33 $\%$} & \multicolumn{2}{c|}{1.32 $\%$ }\\
\hline
\end{tabular}
\end{table}

We have assumed a Poisson distribution to model the observations but, as seen in Figure~\ref{Fig:mediavarianza}, there is a clear overdispersion in the data set. To account for this overdispersion, the Poisson distribution may be replaced by two alternatives included in the \textbf{hhh4} function: 'NegBin1', that is a Negative Binomial model with a common overdispersion parameter $\psi$ for all areas and 'NegBinM'  that has different overdispersion parameters ($\psi_i$) for the different health areas, but these distributions do not improve the fit (see Table~\ref{tabla:seleccion modelos}).

\begin{table}[H]
\caption{Goodness of fit comparison between Poisson and two Negative Binomial distributions.\label{tabla:seleccion modelos}}
\begin{tabular}{|c|ccc|}
\hline
    &Poisson  &NegBin1 & NegBinM\\
\hline
Log-likelihood: & -8539& & \\
AIC: & 17177.99& 31554.37	&  31309.46\\
BIC: &17488.74 &31566.80 &31464.84\\
\hline
\end{tabular}
\end{table}

This  basic model can be refined. Paul et al.
\cite{paul2011predictive} introduced random effects for endemic-epidemic models, which are useful if the areas exhibit heterogeneous incidence levels not explained by observed covariates allowing for area-specific intercepts in the endemic or epidemic components.  Due to the heterogeneity shown in the different health departments, Model 2 considers an  area-specific baseline incidence, $\alpha^{\nu}_k$; the population fraction $e_k$ has been  included as a multiplicative offset and $\alpha^{\phi}_k$ reflects the mean spatial force of influence of the neighbouring health areas.

Both  $\alpha^{\nu}_k$ and $\alpha^{\phi}_k$ have been modelled as fixed effects.

\begin{itemize}
\item Model 2:
\begin{eqnarray}\label{hhh4Model1}
\mu_{kt}&=& e_{k}\nu_{kt} +\lambda_{kt} Y_{k, t-1}+\phi_{kt} \sum_{q\neq k} w_{qk} Y_{q,t-1}, \ \ \ \ \nu_{kt}, \lambda_k, \phi_k >0,\nonumber\\
log(\nu_{kt})   &=& \alpha^{\nu}_k+\beta_1^{\nu} x_{kt}, \hbox{     }\forall k \nonumber \\
log(\lambda_{kt})   &=& \alpha^{\lambda}, \hbox{     }\forall t,k  \nonumber \\
log(\phi_{kt})     &=& \alpha^{\phi}_k, \forall t \nonumber\\
w_{qk} &=&  I(q \sim k).
\end{eqnarray}
\end{itemize}

So, considering a Poisson distribution for the observations, the goodness of fit measures of Model 2 and the contribution of the three components of the mean to the global fit are shown in Table~\ref{tabla:parametrosmodelo2}. Once again, results are shown considering the smoothed number of positive new cases at a lag of 9 (Model 2.1) and the smoothed number of new positive cases at lags of 9 and  5 (Model 2.2)  as covariates. Individual estimates for the parameters are not shown here, because of the quantity of parameters to estimate.

\begin{table}[H]
\caption{Goodness of fit measures and contribution of the endemic and autoregressive components to the global fit. RMSEf is the RMSE of fitted values and RMSEp the RMSE of predictions.}\label{tabla:parametrosmodelo2}
\begin{tabular}{|c|c|c|}
\hline
& Model 2.1 & Model 2.2 \\
\hline
Log-likelihood: & -8539  & -8538.49\\
AIC: &   17177.99& 17178.98 \\
BIC: &   17488.74& 17495.94 \\
\hline
RMSEf & 0.71 & 0.69\\
RMSEp & 5.76& 5.78\\
\hline
endemic &   1.89 $\%$ & 2.04 $\%$  \\
epi.own &   94.73 $\%$ & 94.60 $\%$\\
epi.neigbours &  3.38 $\%$ & 3.36 $\%$ \\
\hline
\end{tabular}
\end{table}

As can be seen in Tables~\ref{tabla:parametrosmodelos1} and \ref{tabla:parametrosmodelo2}, there is not much difference between the four models.  In all of them, the largest portion of the fitted mean results from the within-area autoregressive component  (between 94 and 97$\%$), with very little contribution of cases from adjacent
areas and a rather small endemic incidence. There are also no great differences in the goodness of fit parameters provided by the Likelihood inference (Log-likelihood, Akaike Information Criteria (AIC), Bayesian Information Criteria (BIC)) or in the values of the RMSE.

\begin{figure}
\includegraphics[scale=0.5]{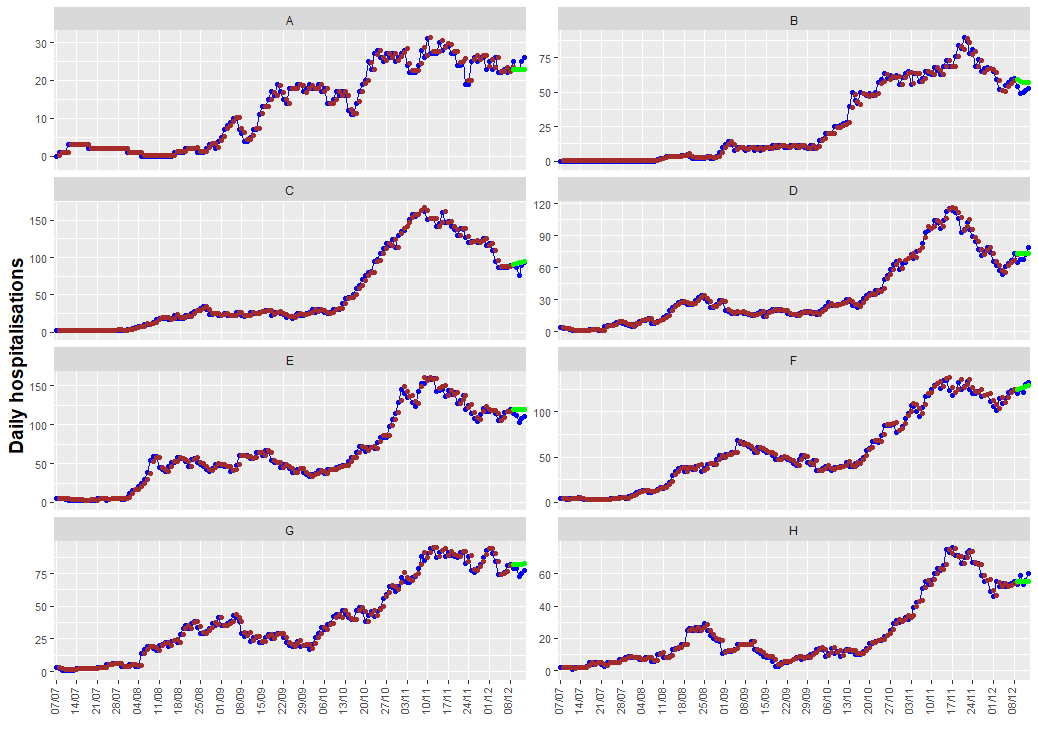}
\caption{Observed values (in blue) jointly with fitted values (in brown) and predictions (in green).\label{Fig:hhh4modelo1}}
\end{figure}

Figure~\ref{Fig:hhh4modelo1} shows up to 5-day predictions  obtained with Model 2.2, together with the fitted and the  true observed values.

\subsection{Multivariate Covariance Generalised Linear Models}  \label{Appl:MCGLM}

In this section, we apply the MCGLM approach to analyse the multivariate count data set that was presented in Section \ref{Sec:Dataset}, Eq. \ref{eq:mcglm1}.

As seen in Eq. \ref{eq:mcglm1}, MCGLM takes non-normality into account,
defining a variance function and modelling the mean structure
by means of a link function and a linear predictor.
In this application, we have daily observations from  24 health departments ($K=24$) on $N=533$ consecutive days.

Despite the previous model, we are no longer dealing with an autoregressive model, but this autoregressive effect can be included in the linear predictor expression together with other covariates

\begin{eqnarray}\label{Eq:mediamcglm}
g(\mu_{kt})= \beta_0+\beta_1 x_{kt} +\beta_2 Y_{k, t-1}+\beta_3 \sum_{q\neq k} w_{qk} Y_{q,t-1}+e_{kt}.
\end{eqnarray}
with $g()$  being, in this case, the log-link function and $e_{kt}$ the offset.

A key point for specifying McGLMs for mixed types of outcomes is the specification of the covariance
matrix within outcomes $\Sigma_k$. Following \cite{bonat2016multivariate}, we define the covariance within outcomes by
\begin{eqnarray}
\Sigma_t = V(\mu_t; p_t)^{\frac{1}{2}}(\Omega(\tau_t))V(\mu_t; p_t)^{\frac{1}{2}}.
\end{eqnarray}

The matrix linear predictor is defined as: $ h(\Omega(\tau_t))=\tau_{0}I_{n\times n}+\tau_1 Z_1+\tau_{2}Z_2$,
where $n$ denotes  the total number of observations in the data set  and $I_{n\times n}$ is the $n\times n$ identity matrix. $\tau_{0}$ is the intercept of the covariance linear model. If $n_t$ denotes the number of observations in time in each spatial region,  $Z_1=I_{24\times 24}\bigotimes_G \Gamma_t$, where  $\bigotimes_G$ denotes the Kronecker product and $\Gamma_t=(\gamma(i,j))_{i,j \in\{1,\cdots,n_t\}}$ with $\gamma(i,j)=1 $ if $j\in \{i-1,i+1\}$ and $0$ otherwise. $\tau_1$ measures the effect of the 'time'. Finally,
$Z_2=W \bigotimes_G I_{n_t \times n_t}$ with $W$ being the spatial adjacency matrix between the 24 health departments.

In this case, we have obtained the best fits when using the exponential covariance link function as $h()$.

To define $V(\mu_t; p_t)$, we will adopt the  Poisson-Tweedie dispersion function  \cite{jorgensen2016discrete} so that
\begin{eqnarray}
V(\mu_t; p_t)=diag(\mu_t^{p_t})
\end{eqnarray}
in accordance with the plot in Figure~\ref{Fig:mediavarianza}.


We employed a step-wise procedure for selecting the components of the linear predictor. As in the previous modelling, we are going to consider the number of new positive cases at a lag of 9, the number of new positive cases at a lag of 5 and the number of hospitalisations at lag of 1 as potential covariates in Eq. \ref{Eq:mediamcglm}, and an additional categorical covariate \textit{'Health Department'} to allow different intercepts ($\beta_{0k}$ instead of $\beta_0$) in Eq. \ref{Eq:mediamcglm}. The population of each health area will also be used as an offset. The SIC using penalty $\delta = 2$ and the Wald test were used in the forward and backward steps, respectively. We defined a stopping criterion for the selection procedure as SIC $>$ 0, since the penalty is larger than the score statistics in that case.

The SIC is an important tool to assist with the selection of the linear and matrix linear predictor components, but it is less useful for comparing models fitted using different link, variance or covariance functions. The mcglm package implements the SIC to select the linear and matrix linear predictor components, with the mc\_sic and mc\_sic\_covariance functions.
In this case, the SIC values indicate that all components, except the value of the daily hospitalisations with a lag of 1 in neighbouring regions, should be included in the model (SIC $<$ 0).

To compare the goodness of non-nested models, the \textbf{gof} function provides the pseudo Akaike information criterion (pAIC), the pseudo Bayesian information criterion (pBIC) and the pseudo Kullback-Leibler information criterion (pKLIC). We are going to use it to select the best structure for the matrix linear predictor. The results can be seen in Table \ref{tabla:goodness_mcglm}. We will fit the model considering $h(\Omega(\tau_t))=\tau_{0}I_{K\times K}+\tau_1 Z_1+\tau_{2}Z_2$.


\begin{table}[H]
\caption{Goodness of non-nested models, defined from different matrix linear predictors $h(\Omega(\tau_t))$.\label{tabla:goodness_mcglm}}
\begin{tabular}{|c|c|c|c|}
\hline
&\multicolumn{3}{c|}{$h(\Omega(\tau_t))=$}  \\
\hline
& $\tau_{0}I_{K\times K}$ &$\tau_{0}I_{K\times K}+\tau_1 Z_1$ &
$\tau_{0}I_{K\times K}+\tau_1 Z_1+\tau_{2}Z_2$ \\
\hline
 plogLik &-13540.43 &-10448.25  &-10444.49\\
 df  & 28&29 &30 \\
 pAIC & 27136.86 & 20954.5     &20948.98\\
 pKLIC &27138.46 &20958.28      &20954.1\\
 pBIC&27311.06 & 21134.92&21135.62\\
\hline
 RMSEf& 13.72 & 13.62& 11.39\\
 RMSEp& 16.46 &11.77 &11.6\\
\hline
\end{tabular}
\end{table}

\begin{figure}
\includegraphics[scale=0.7]{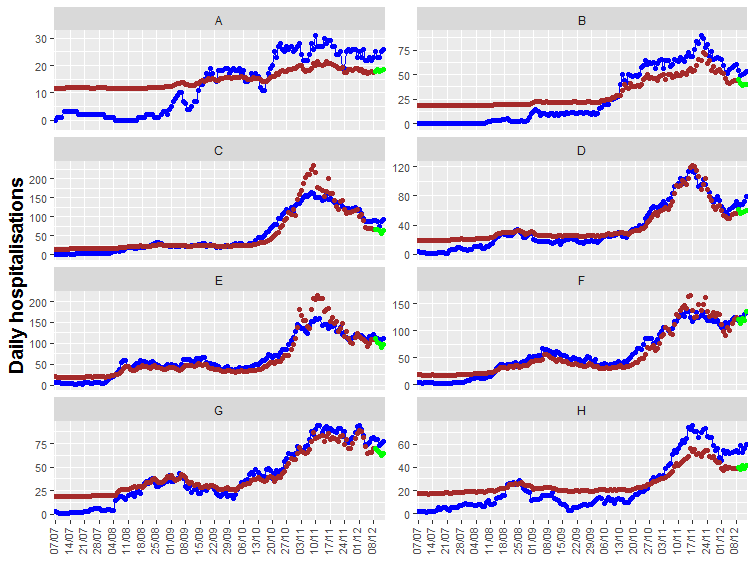}
\caption{Observed values (in blue) together with fitted values (in brown) and predictions (in green).} \label{Fig:mcglmfit}
\end{figure}

Figure~\ref{Fig:mcglmfit} shows the up to 5-day predictions  obtained with the resulting model, together with the fitted and the  true observed values. In this case, we have obtained an RMSEp equal to 11.6, quite higher than the obtained with the other models/packages. Fig. \ref{Fig:mcglmfit} shows that in this case there are health departments where the model provides neither a good fit of the observations nor precise predictions, while the fits and predictions of other health  departments are very accurate. With the information available, we have not found a model with better fits and predictions for all the health departments.

\subsection{Bayesian spatiotemporal models} \label{Appl:Bayes}

In this section, we apply different models included in the CARBayesST package to the COVID-19 data given in Section  \ref{Sec:Dataset}.
For the reasons explained in this section, we use the Poisson log-linear model for $Y_{kt}$, Eq. \ref{eq:hierarchical1}. Overdispersion cannot be controlled with this package. Although that could be regarded as a handicap, as we will see in the results sections, the adjusted and forecasted values are quite good.

In order to induce spatial smoothness between the random effects, we use the binary adjacency matrix $W$ used in previous sections. Element $w_{ik}$ =1 if areas $i$ and $k$ share a common border and $w_{ik}$=0 otherwise, whereas $w_{ii}$=0 for all $i$.

Taking into account again the characteristics of our data, the plots shown in Section \ref{Sec:Dataset} and the fact that our main objective is the prediction of future observations, just two of the  spatiotemporal correlation structures included in this package (see Table~\ref{TablaBayesianModels}) have been fitted  for $\psi_{kt}$ in Eq. \ref{eq:hierarchical2}: the CARar and CAR adaptative structures.
Neither spatially varying linear time trend models nor spatiotemporal clustering models are appropriate for our data and CARanova and CARsepspatial structures assume a symmetric temporal correlation that does not allow us to obtain future predictions.



In both cases, the spatiotemporal structure is modelled with a multivariate first-order autoregressive process with a spatially correlated precision matrix:
\begin{eqnarray}
  \psi_{kt}      & = & \phi_{kt} \\ \label{CARadaptative}
  \phi_{t}|\phi_{t-1} & \sim & N(\rho_T \phi_{t-1}, \tau^2Q(W,\rho_S)^{-1})\\ \nonumber
  \phi_{1}& \sim & N(0, \tau^2 Q(W,\rho_S)^{-1}), \\\nonumber
  \tau^2 & \sim & Inverse-Gamma(a,b),\\\nonumber
  \rho_S, \rho_T & \sim & Uniform(0,1),\nonumber
\end{eqnarray}
$\phi_t=(\phi_{1t},\ldots,\phi_{Kt})$ is the vector of random effects for time period $t$, the precision $Q(W,\rho_S)$ corresponds to the CAR models proposed in  \cite{leroux2000estimation} and has the expression:
$$Q(W,\rho_S)=\rho_S(diag(W \textbf{1})-W)+(1-\rho_S)I,$$
$\mathbf{1}$ is the $K\times1$ vector of ones and $I$ the $K \times K$ identity matrix.
$( \rho_S , \rho_T )$ respectively control the levels of spatial and temporal autocorrelation, with values of 0 corresponding to independence while a value of 1 corresponds to strong autocorrelation.

The random effects from CARar have a single level of spatial dependence that is controlled by the parameter $\rho_S$. That means that all pairs of adjacent areal units will have the same degree of autocorrelation: strongly if $\rho_S$ is close to one, while no spatial dependence will exist if $\rho_S$ is close to zero.

The CARadaptative model allows for localised spatial autocorrelation, that is, it allows it to be stronger in some parts of the study region. This could be adequate for our data because it would be possible for spatial autocorrelation between adjacent health departments to be correlated or conditionally independent, depending, for example, on whether these departments are in the same big city or according to the socioeconomic characteristics of their inhabitants.

The CARadaptative model allows this spatial autocorrelation heterogeneity by allowing spatially neighbouring random effects, which is achieved by modelling the non-zero elements of the neighbourhood matrix $W$ as unknown parameters rather than assuming they are fixed constants.

With respect to the covariates to be included in the mean,  for the same reasons explained in the previous sections, we again tried two possibilities. Case 1: the  smoothed number of new positive cases at a lag of 9. Case 2: the  smoothed number of new positive cases at a lag of 9 plus the smoothed number of new positive cases at a lag of 5.

Additionally, as a basic area-specific measure of disease incidence, the population fraction $e_k$ has been also included as an offset.


Inference for all models is based on thinning (by 10) 60,000 posterior samples, including a burn-in period of a further 1000 samples. Convergence plots assured that it was reached in all cases.

The Table~\ref{tabla:seleccion modelos Bayesianos} displays the overall fit of each model by presenting the deviance information criterion
(DIC) and the effective number of parameters (pd). It shows that the adaptive model fits the data better than the pure AR model, with reductions in the DIC in both cases.
As a complementary measure of goodness of fit and in order to compare across models, the mean square error of adjusted data  (RMSEf) has been also calculated and it is showed  in Table~\ref{tabla:seleccion modelos Bayesianos}. In this case better results are obtained with the CARar model but as it can be seen the differences are very slight.

\begin{table}[H]
\caption{Deviance information criterion (DIC), effective number of parameters (pd), RMSE of fitted values (RMSEf) and RMSE of predictions (RMSEp) for each model and scenario.\label{tabla:seleccion modelos Bayesianos}}
\begin{tabular}{|c|c|c|c|c|}
\hline
    &\multicolumn{2}{c|}{$CARar$}  &\multicolumn{2}{c|}{$CARadaptative$} \\
\hline
    &case 1 & case 2 & case 1 & case 2\\
\hline
DIC & 18730.92&  18741.31  &  18671.93  &  18678.58   \\
pd&793.5918 &  787.9688  & 780.8599 &770.0481 \\
\hline
RMSEf& 1.939859&2.027429 & 1.970366 & 2.059419 \\
\hline
RMSEp& 5.610704&5.854628 & 5.861029 & 5.402237 \\
\hline
\end{tabular}
\end{table}

The medians of the posterior distribution of each parameter and its $95\%$ credible intervals are displayed in Table~\ref{tabla:par modelos Bayesianos}.

As can be seen, the estimated parameters of spatial and temporal correlations show a strong spatial and temporal correlation in all cases. Regarding the covariates, both are significative.

Finally,  with respect to the number of step-changes between two spatially adjacent areas detected in the CARadaptative models, only one step change is detected in case 1, while no changes are detected in case 2.

\begin{figure}

\includegraphics[scale=0.85]{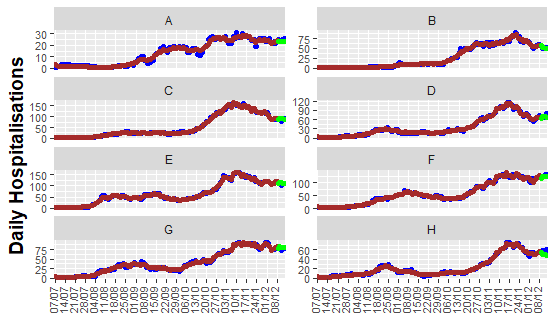}
 \caption{ Fitted and predicted values in brown and green together with the observed counts in blue.\label{Fig:PredBayes}}

\end{figure}

 Since our main objective is prediction, we calculate the mean square error of the prediction for up to 5 days for the four cases. These values can be found in the last row of Table~\ref{tabla:seleccion modelos Bayesianos}, and again no great differences are observed. In this case, the best result is obtained with the CARadaptative model with both covariates: positive cases at lags of 9 and 5. Adjusted, observed and predicted values of these models can be seen in Figure~\ref{Fig:PredBayes}.

\begin{table}[H]
\caption{Medians of the posterior distribution of each parameter and 95 \% credible intervals for each model and scenario\label{tabla:par modelos Bayesianos}}
\scalebox{0.78}{
\begin{tabular}{|c|ccc|ccc|ccc|ccc|}
\hline
&\multicolumn{3}{c|}{CARar case 1}&\multicolumn{3}{c|}{CARar case 2}&\multicolumn{3}{c}{CARadaptative case 1}&\multicolumn{3}{c|}{CARadaptative case 2}\\
\hline
&Median & 2.5 \% & 97.5\% &Median  & 2.5 \% & 97.5\%&Median  & 2.5 \% & 97.5\%&Median  & 2.5 \% & 97.5\%\\
\hline
Intercept&5.7011 &5.5108 &5.7307  &5.6233& 5.2602 &5.6621& 5.6955 & 5.6223 &  5.7247  & 5.6107 &  5.3673  & 5.6532 \\
Posit9 &0.0022 &0.0011 &0.0108&0.0027& 0.0018& 0.0100& 0.0022 & 0.0012  & 0.0054 & 0.0028  & 0.0018 &  0.0081   \\
Posit5&-&-&-&0.0025 &0.0014& 0.0097& -&-&-&0.0026 &  0.0017 &  0.0075\\
tau2& 0.0225& 0.0197& 0.0313&0.0222& 0.0193& 0.0313 & 0.0136 & 0.0099  & 0.0206 &  0.0128 &  0.0094  & 0.0198  \\
rho.S&0.9432 &0.8767 & 0.9591 & 0.9666 & 0.9349 &  0.9784 & 0.9652 &  0.9225  & 0.9774 & 0.9671  & 0.9206  & 0.9792  \\
rho.T&0.9895 &0.9811 &0.9949  &0.9376& 0.8420 &0.9548&   0.9898 & 0.9836 &  0.9952& 0.9890 &  0.9824 &  0.9946\\
tau2.w&-&-&-&-&-&-&  171.0751 & 98.3797 & 289.3351  & 177.6793 & 104.1783 & 293.4227 \\
\hline
\end{tabular}}
\end{table}

\section{Discussion }\label{Discussion}
Having reviewed and applied the different models, it has been seen that the second model (mcglm package) is the most flexible for modelling the variable of interest, $Y_ {kt}$, allowing any type of function of time $t$ in the expression of the mean and a huge variety of spatiotemporal neighbourhoods to model both the mean and the variance-covariance matrix (when necessary). The first model (surveillance package) can also use any function of time in the expression of the mean, but the temporal relationship with the neighbours only allows an autoregressive (AR) structure of order 1. This model allows us to choose between a Poisson distribution for modelling the observations or a Negative Binomial model when the data show overdispersion. Regarding modelling the mean as a function of $t$, the third model (CARBayesST package) only offers the possibility of linear relationships. In this third model, spatiotemporal relationships between the observations can be introduced into the model of the mean using random effects with different correlation structures.

Having fitted the models, the surveillance package provides the oneStepAhead  function to compute successive one-step-ahead predictions for the fitted model, in addition to confident intervals for the predictions and plot methods.  However, the other two packages do not have  any function implemented to predict future observations from the fitted models, making this process more difficult for non-experts in R programming.

Regarding the estimates of the parameters of the models, significant positive parameters for the covariates are obtained in all cases and the parameters that indicate temporal correlation show high values too for all models. Regarding spatial  autocorrelation, there is a difference between the models depending on how the spatial neighbourhood has been included in the model. In general, it can be considered that in order to predict the number of hospitalisations per day, both the number of hospitalisations from the previous day and the number of new cases in the region of interest and in adjacent areas are needed.

Because each type of model has a different methodology that provides different measures of goodness of fit,  to compare the performance across different approaches two common and habitual measures are calculated; the RMSE of the predictions up to 5 days,  in order to compare prediction performance between different models and the RMSE of adjusted data as a measure of goodness of fit.

With respect to the results, the RMSE of the predictions up to five days of the best model in each package ranges from 5.4 (using the CARBayesST package) to 13.72 (using the mcglm package). There are no large differences between the fitted models using CARBayesST and Surveillance packages, ranging in this case from 5.4 to 5.78, both values are acceptable in clinical practice. RMSE of fitted data are too excellent using CARBayesST and Surveillance (from 0.53 to 1.93),  obtaining in this case the smallest error with the Endemic-Epidemic model \ref{hhh4Model0} (case 2).

As was previously explained using Multivariate Covariance generalised Linear Models there are health departments where the model provides neither a good fit of the observations nor precise predictions, while the fits and predictions of other health departments are very accurate. With the information available, we have not found a model with better results for all the health departments. But, as has also been said before, this is the most flexible model for modelling spatiotemporal count data response variables and, probably, good results could have been obtained if there had been more information or in other types of applications.

These models can  truly be used in the current situation. As far as we know, the surveillance package has been used to address several problems related to the COVD-19 pandemic. In fact, the endemic-epidemic models included in the surveillance package, have been applied to a multitude of infectious diseases (see \cite{dunbar2020endemic} for references) such as Influenza \cite{paul2008multivariate}, Norovirus \cite{held2017probabilistic} and COVID-19 \cite{fronterre2020covid,giuliani2020modelling}.   In fact, a regularly updated table of use cases is maintained by S. Meyer at \url{https://github.com/rforge/surveillance/blob/master/www/applications_EE.csv}. The CARBayesST package has also been used recently for the study of case-fatality risk due to COVID-19 in Colombia \cite{polo2020bayesian}.

From the beginning of the 2020, other works have been also focused on modeling the number of COVID-19 related hospitalizations, but as far as we know, all them have objectives, covariates and methodologies different from those seen in our work.  Ferstad et al. \cite{ferstad2020model} model the number of people in each county in the United States who are likely to require hospitalization as a result of COVID-19 given the age distribution of the county per the US Census.  G. Perone \cite{perone2020comparison} compares several time series forecasting methods to predict the  number of patients hospitalized with mild symptoms, and in intensive care units (ICU) in Italy, over the period after October 13, 2020, getting RMSE values greater than ours. Reno et al. \cite{reno2020forecasting} model the spread of COVID-19 and its burden on hospital care under different conditions of social distancing in Lombardy and Emilia-Romagna, the two regions of Italy most affected by the epidemic, using a Susceptible-Exposed-Infectious-Recovered (SEIR) deterministic model, which encompasses compartments relevant to public health interventions such as quarantine. Goic et al. \cite{goic2021covid} combine autoregressive, machine learning and epidemiological models to provide a short-term forecast of ICU utilization at the regional level in Chile.

   Most of them do not provide a goodness of fit measurement that can be used to compare with our results. Just \cite{ferstad2020model} and \cite{goic2021covid} give values of RMSE. These values are in general greater than the obtained in our work, but they are not directly comparable, because their objective is to predict the number of patients hospitalized with mild symptoms, and/or in intensive care units (ICU) separately.

Within the framework of the government support group of the Generalitat Valenciana our models are intended to be used and updated weekly, helping the
government make public health decisions such as, the possible need to open new COVID-19 wards in hospitals in the most affected regions.

\section{Conclusions }\label{conclusions}
The aim of this work is to review three different spatiotemporal models for count data, implemented in R packages, and to test their performance on an actual case study using three completely different approaches.

Due to these different statistical methodologies, the different packages provide different goodness of fit  measures, there being no measure in common between them. Therefore, as the final aim of our case study has been the short-term prediction of the evolution of hospitalisations in the different spatial areas, the root mean squared prediction error (RMSE) has been obtained in all cases.
We can achieve very satsifactory results using each of the packages reviewed and there is not much difference between them. These results are very promising in the particular case of the Valencian Community, but they are also very valuable because they can be applied to any region and the use of these models  can be promoted to help in short-term government decision making regarding  preventive measures against the collapse of hospitals. Additionally, they can provide tools to know in advance whether it is necessary to expand hospital capacity in terms of beds and/or workers.

Our objective in this work has not been to say that one package or one type of model is better than others, but to show possibilities that can be used in practice to analyse this type of data. The choice of the type of model to use will depend on the application at hand.

\section*{Acknowledgements}
This research is supported by Ayudas Fundación BBVA a Equipos de Investigación Científica SARS-CoV-2 y COVID-19 and FONDO SUPERA COVID19 by Banco Santander. We would like to thank to Dr. Nuria Oliver (chief head of Generalitat Valenciana Task force) and Generalitat Valenciana for their support providing the data set.

\end{document}